\documentclass[a4paper]{jpconf}
\usepackage{graphicx}
\usepackage{units}
\usepackage[utf8]{inputenc}

\newcommand{\fauthor}[3]{#2$^{#1}$#3}
\newcommand{\faddress}[2]{\address{$^{#1}$ #2}}

\usepackage{multicol}
\usepackage{etoolbox}
\usepackage{relsize}

\begin{document}
\patchcmd{\thebibliography}
  {\list}
  {\begin{multicols}{2}\smaller\list}
  {}
  {}
\appto{\endthebibliography}{\end{multicols}}

\title{Results of the first NaI scintillating calorimeter prototypes by COSINUS}

\author
{
  \fauthor{1,*,}{F Reindl}{\footnote[9]{Now at affiliations $^{5,6}$}}
 \fauthor{2}{G Angloher}{,}
 \fauthor{3,4}{P Carniti}{,}
 \fauthor{3,4}{L Cassina}{,}
 \fauthor{3,4}{L Gironi}{,}
 \fauthor{3,4}{C Gotti}{,}
 \fauthor{5,6}{A G\"utlein}{,}
 \fauthor{2}{M Maino}{,}
 \fauthor{2}{M Mancuso}{,}
 \fauthor{8}{N Di Marco}{,}
 \fauthor{7,8}{L Pagnanini}{,}
 \fauthor{3}{G Pessina}{,}
 \fauthor{2}{F Petricca,}{,}
 \fauthor{8}{S Pirro}{,}
 \fauthor{2}{F Pr\"obst}{,}
 \fauthor{5,6}{R Puig}{,}
 \fauthor{7}{K.~Sch\"affner}{ and }  
 \fauthor{5,6}{J Schieck}{}
}
%\fauthor[a,1]{F.~Reindl \note{corresponding author, present address: INFN, Sezione di Roma 1, I-00185 Roma - Italy},}
%\emailAdd{florian.reindl@mpp.mpg.de}
%\thanks{present address: INFN, Sezione di Roma 1, I-00185 Roma - Italy}
%\mpi

%\emailAdd{karoline.schaeffner@lngs.infn.it}
%\fauthor[a,3]{W.~Seidel \note{deceased 19/02/2017}}

\faddress{1}{INFN - Sezione di Roma 1, I-00185 Roma - Italy}

\faddress{2}{Max-Planck-Institut f\"ur Physik, D-80805 M\"unchen - Germany} 

\faddress{3}{INFN - Sezione di Milano Bicocca, I-20126  Milano - Italy}

\faddress{4}{Dipartimento di Fisica, Universit\`{a} di Milano-Bicocca, I-20126 Milano - Italy}

\faddress{5}{Institut f\"ur Hochenergiephysik der \"OAW, A-1050 Wien - Austria}

\faddress{6}{Atominstitut, Technical University Vienna, A-1020 Wien - Austria}

\faddress{7}{Gran Sasso Science Institute, I-67100 L'Aquila - Italy}

\faddress{8}{INFN - Laboratori Nazionali del Gran Sasso, I-67010 Assergi - Italy}

%%%%%%%%%%%%%%%%%%%%%%%%%%%%%%%%%%%%%%%%%%%%%%%%%%%%%%%%%%%%%%%%%%%
%\affiliation[]{Dipartimento di Scienze Fisiche e Chimiche - Universit\`{a} degli studi dell'Aquila, I-67100 Coppito - Italy}}

\ead{florian.reindl@oeaw.ac.at}
\setcounter{footnote}{0}
\begin{abstract}
  Over almost three decades the TAUP conference has seen a remarkable momentum gain in direct dark matter search. An important accelerator were first indications for a modulating signal rate in the DAMA/NaI experiment (today DAMA/LIBRA) reported in 1997. Today the presence of an annual modulation observed by DAMA, which matches in period and phase the expectation for dark matter, is doubtless and supported at $>9\sigma$ confidence. Despite the positive evidence from the DAMA experiment the underlying nature of dark matter is still considered an open and fundamental question of nowadays particle physics. No other direct dark matter search experiment could confirm the DAMA claim up to now; moreover, numerous null-results are in clear contradiction with DAMA under so-called standard assumptions for the dark matter halo and the interaction mechanism of dark with ordinary matter. As both bear a dependence on the target material, resolving this controversial situation will convincingly only be possible with an experiment using sodium iodide (NaI) as target, just like DAMA. COSINUS aims to even go a step further by combining NaI with a novel detection approach. DAMA and all other NaI experiments solely measure the scintillation light created by a particle interaction in the NaI crystal. COSINUS aims to operate NaI as a cryogenic calorimeter reading scintillation light and phonon/heat signal. Two distinct advantages arise from this approach, a substantially lower energy threshold for nuclear recoils and particle identification on an event-by-event basis. These key benefits will allow COSINUS to clarify a possible nuclear recoil origin of the DAMA signal with comparatively little exposure of $\mathcal{O}$(100kg days) and, thereby, answer a long-standing question of particle physics. Today COSINUS is in R\&D phase; in this contribution we show results from the 2nd prototype, albeit the first one of the final foreseen detector design. The key finding of this measurement is that pure, undoped NaI is a truly excellent scintillator at low temperatures: We measure 13.1\% of the total deposited energy in the NaI crystal in the form of scintillation light (in the light detector).
 \end{abstract}

\section{Introduction}

Today cosmology precisely measures the dark matter (DM) content of the Universe to exceed the one of ordinary, baryonic matter by a factor of five \cite{adam_planck_2016}, but its underlying nature remains a mystery. Numerous theories for the particle nature of DM were postulated, thereby an all-time favorite are weakly interacting massive particles (WIMPs), thermally produced in the early Universe. Detecting such elusive particles is a severe experimental challenge, in particular in the view of backgrounds any rare-event search has to fight. As proposed in the 1980s DM should lead to an annually modulating rate in the detector due to the earth orbiting around the sun, thus causing a change in its velocity w.r.t the DM halo \cite{drukier_detecting_1986}. As it is practically excluded that backgrounds cause a suchlike modulation, with a phase of one year and peaking in June 2$^{nd}$, observing an annual modulation signal is commonly accepted as a \textit{smoking gun evidence}.

Almost 20 years ago DAMA/NaI (today DAMA/LIBRA) published a first positive indication for such a sought-for modulation \cite{bernabei_searching_1998}. Today the DAMA modulation signal is statistically solid ($>9\sigma$ C.L.) and, furthermore, matches in phase and period the expectation for DM. Other experiments, however, could not yet confirm this observation; in contrary they clearly exclude it in the so-called standard scenario. The latter imposes assumptions on the DM halo as well as on the interaction of DM particles with ordinary matter which both are subject to material dependencies. Thus, only an experiment also based on NaI may bring forward the ultimate = model-independent clarification. Several experiments are committed to this task, ANAIS \cite{coarasa_annual_2017}, COSINE \cite{adhikari_design_2017}, DM-Ice \cite{dm-ice_collaboration_first_2017}, SABRE \cite{froborg_sabre:_2016}, which share with DAMA the principle signal: Scintillation light produced in the Tl-doped NaI target crystals measured with PMTs. In this contribution we discuss results from the COSINUS (Cryogenic Observatory for SIgnals seen in Next-generation Underground Searches) project aiming to operate NaI as cryogenic calorimeter \cite{angloher_cosinus_2016-1}.

\section{A NaI-based Cryogenic Calorimeter}
\begin{figure}
  \begin{minipage}[t]{0.49\textwidth}
   \includegraphics[width=\textwidth]{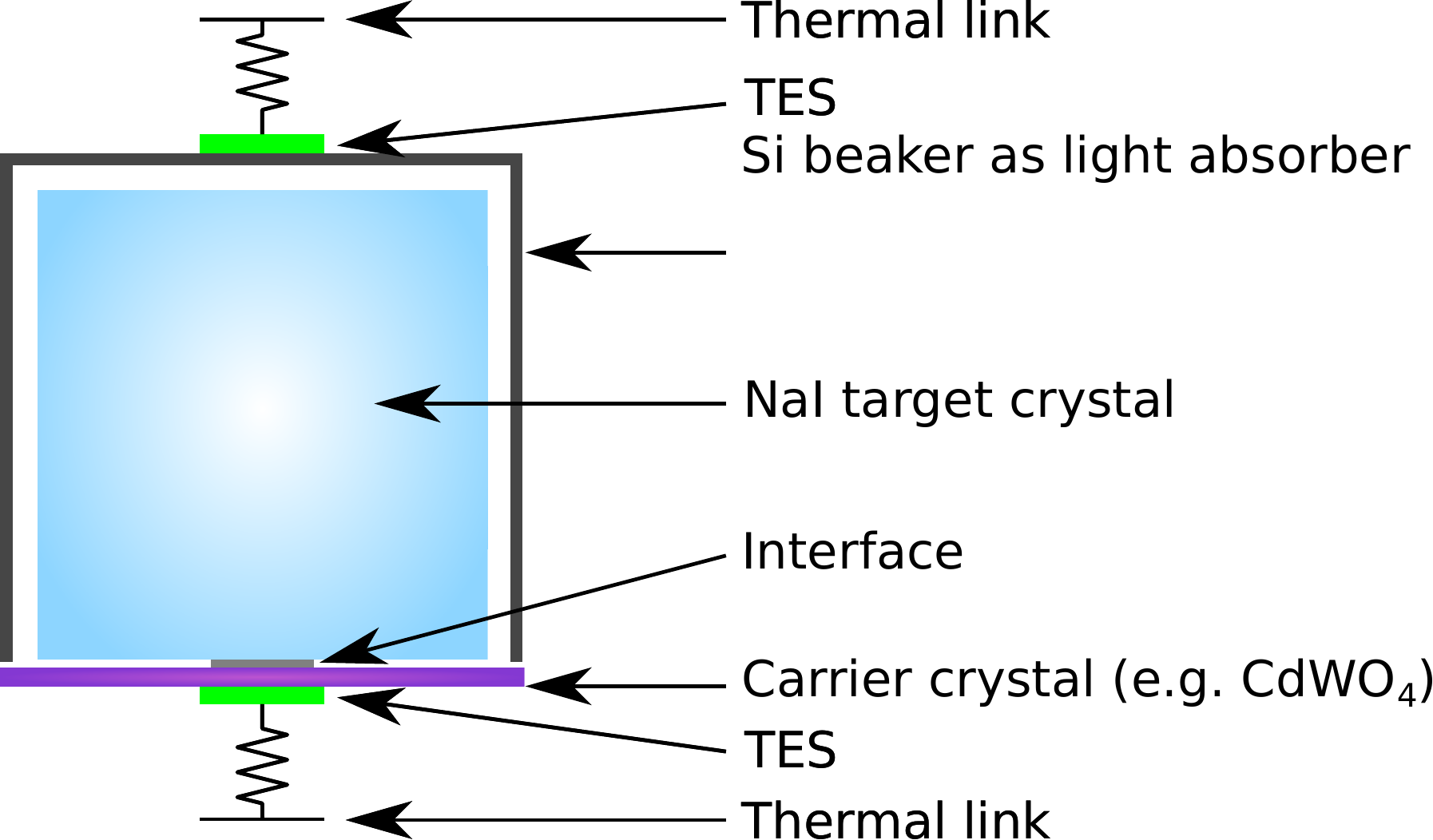}
   \caption{Scheme of a COSINUS detector module. The NaI crystal (blue) had a mass of \unit[66]g, the interface to the carrier crystal (purple) was done with epoxy resin. For more details see text. }
   \label{fig:scheme}
 \end{minipage}%
 \hfill
 \begin{minipage}[t]{0.49\textwidth}
   \centering
  \includegraphics[width=\textwidth]{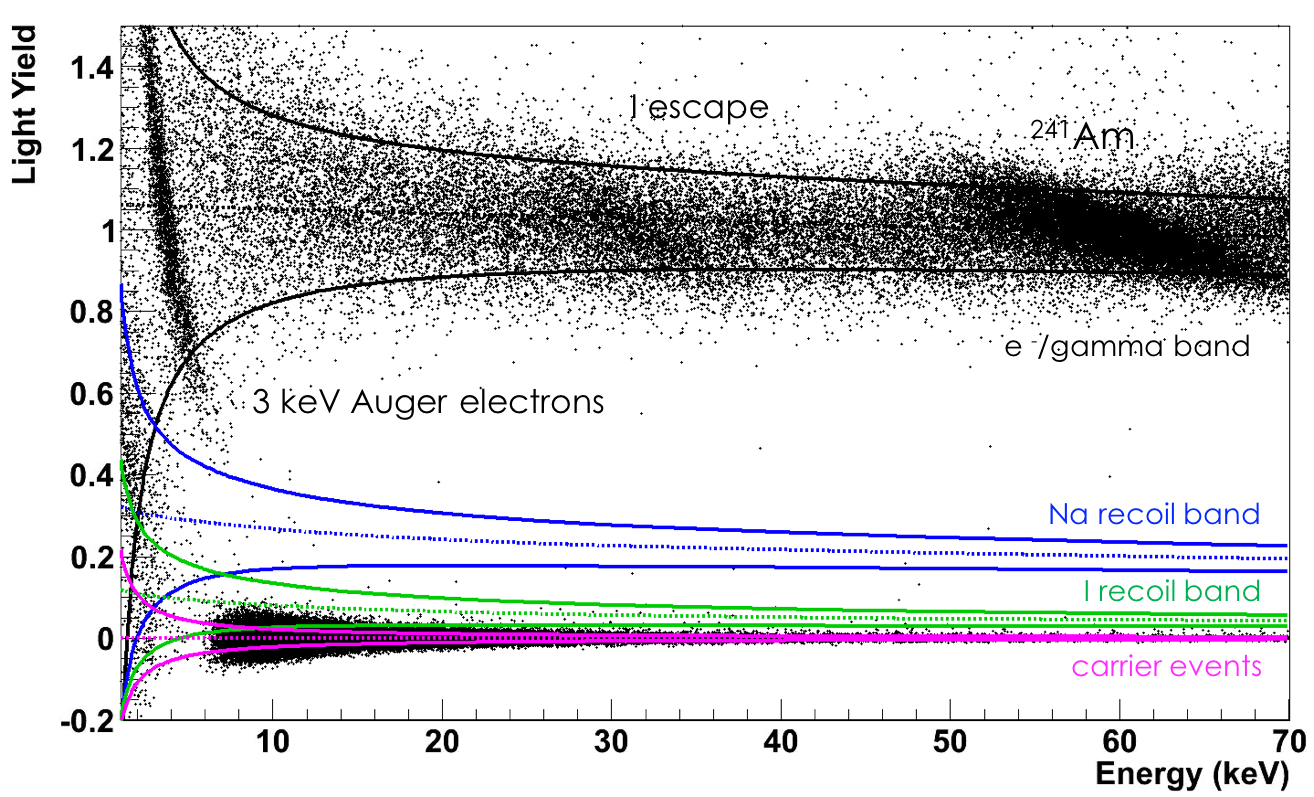}
  \caption{Data from the second COSINUS prototype. Lines mark the central 80~\% bands for electron recoils (black), recoils off Na (blue) and I (green) and events without associated light signal (magenta).}
  \label{fig:LY}
 \end{minipage}
\end{figure}

Cryogenic scintillating calorimeters are successfully used by LUCIFER/CUPID-0 \cite{artusa_enriched_2017} for the search for neutrinoless double-beta decay and by CRESST, the leading experiment for low-mass DM \cite{angloher_results_2016}. Two signals are obtained with such a detector. First, the heat (or phonon) signal which precisely measures the energy deposited in the target crystal, quasi-independent of the interacting particle. Second, the scintillation light which allows for  particle identification on an event-by-event basis, an immensely useful tool to suppress the dominant $\beta/\gamma$-background w.r.t the potential nuclear recoil signal. A scheme of the module design is depicted in figure \ref{fig:scheme}, the central part is the NaI crystal attached to a carrier crystal (CdWO$_4$) \textit{carrying} the Transition Edge temperature Sensor (TES) which reads the heat/phonon signal. The beaker-shaped light detector is drilled out from a high-purity silicon block. It is equipped with a second TES measuring the heat rise of the beaker caused by absorbed scintillation light photons. The ensemble of heat and light detector is called a detector module. The beaker-shape of the light detector not only guarantees a very high light collection efficiency, it also avoids, in combination with the carrier crystal, any line of sight of the NaI crystal to non-active surfaces, thus vetoing backgrounds induced by surface-$\alpha$-decays. Recently the COSINUS collaboration successfully demonstrated for the first time that NaI can be operated as cryogenic calorimeter \cite{angloher_results_2017}. In this article we present results from the follow-up measurement constituting a proof-of-principle test of the final detector design (figure \ref{fig:scheme}). 

\section{Presented Data}

In this contribution we present results obtained from the measurement of the second prototype carried out in the cryogenic test facility of the Max-Planck-Institute for Physics Munich placed at the LNGS underground laboratory in central Italy. The NaI target crystal had a mass of \unit[66]{g} and we collected \unit[1.32]{kg days} of data exposing the detector to $\gamma$-rays (\unit[$\sim$60]{keV}) from an $^{241}$Am-source. We apply certain quality cuts to the data ensuring a stable detector operation and removing events with strongly tilted baselines, pile-up events and in general any event not matching the nominal pulse-shape. 

\section{Pulse-Shape}
\label{sec:pulse-shape}

\begin{figure}
  \begin{minipage}[t]{0.49\textwidth}
   \includegraphics[width=\textwidth]{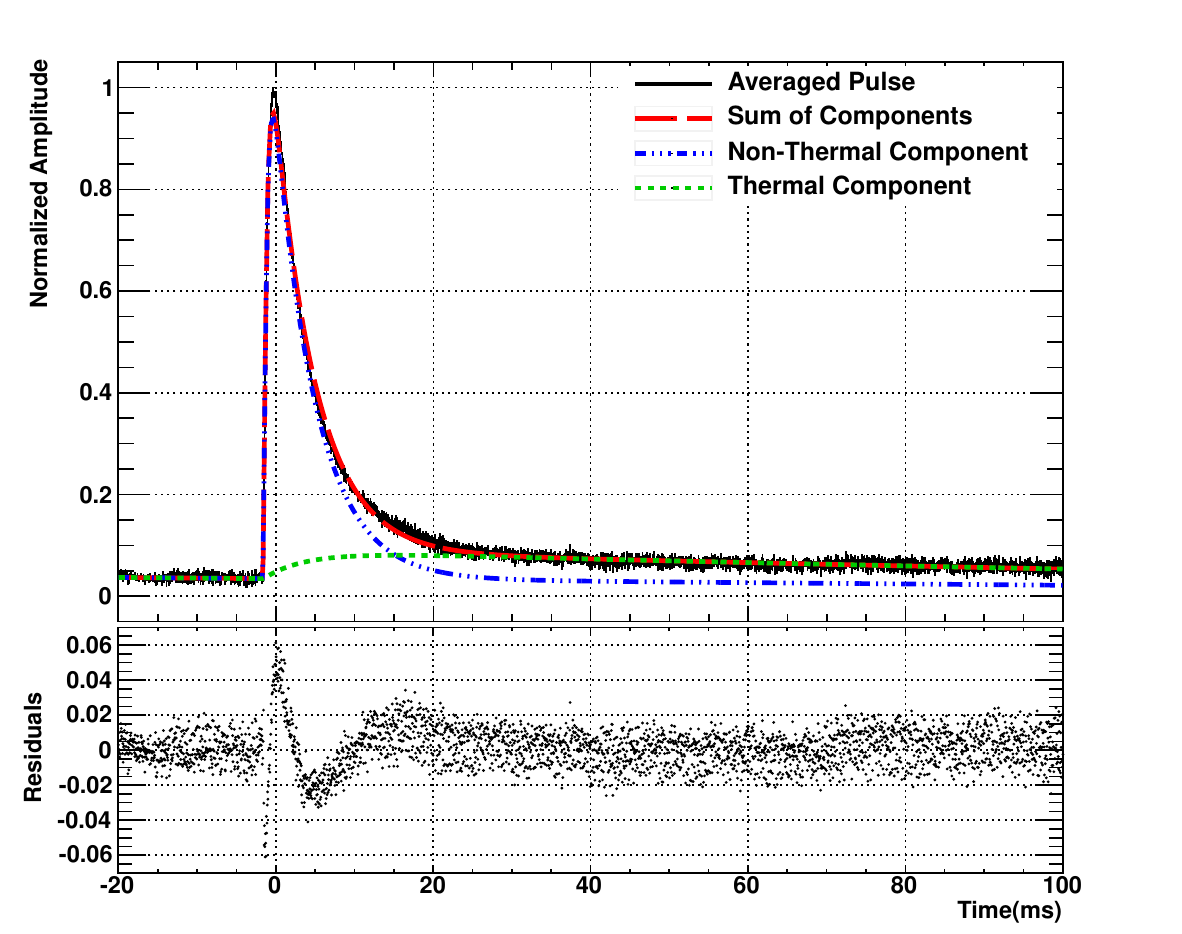}
   \caption{Top: Normalized average pulse (black) and fit of the \textbf{two-component model}. Bottom: Fit residuals.}
   \label{fig:2comp}
 \end{minipage}%
 \hfill
 \begin{minipage}[t]{0.49\textwidth}
   \centering
   \includegraphics[width=\textwidth]{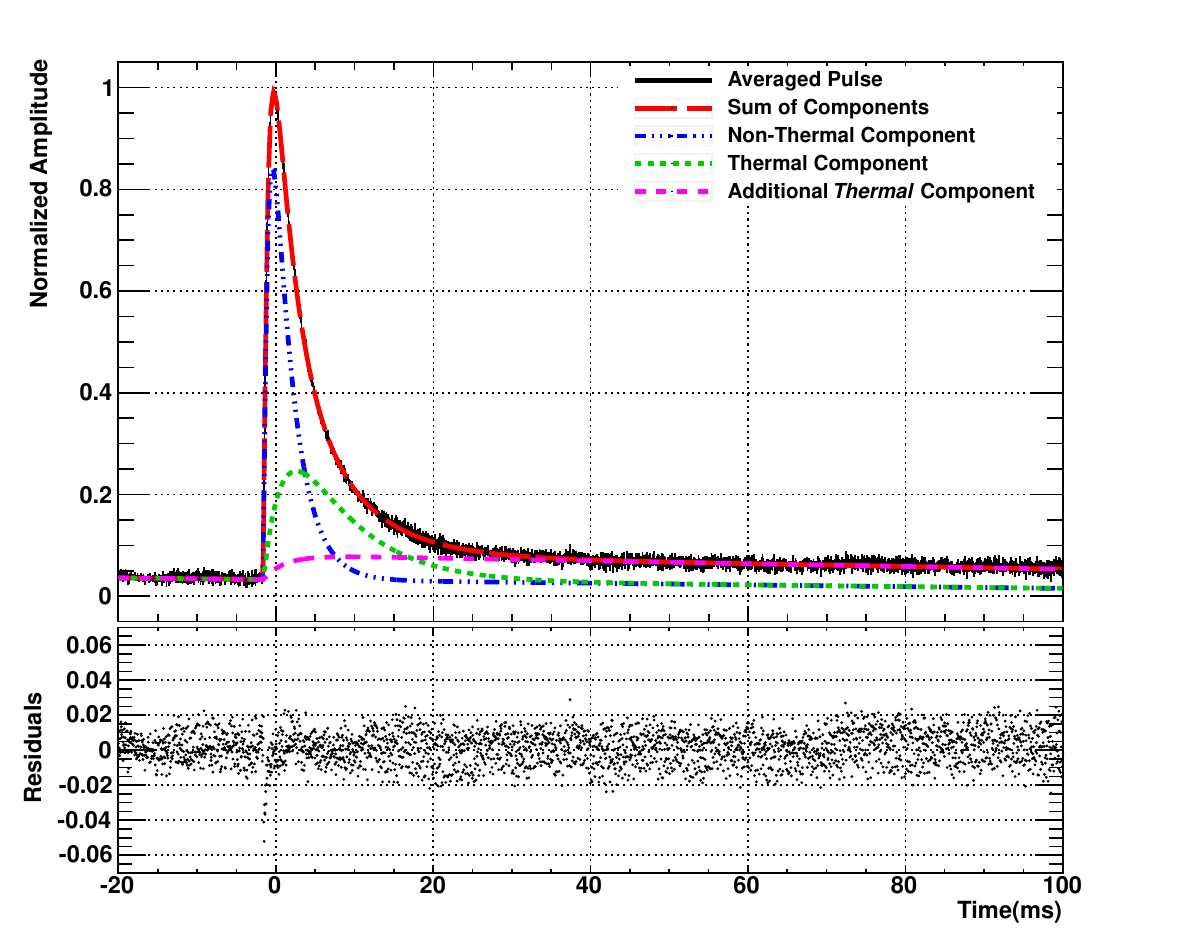}
   \caption{Top: Normalized average pulse (black) and fit of the \textbf{three-component model}. Bottom: Fit residuals.}
   \label{fig:3comp}
 \end{minipage}
\end{figure}

In the measurement of the 1st COSINUS prototype we already found a hint that NaI exhibits a different pulse shape compared to many other materials (Al$_2$O$_3$, CaWO$_4$, ZnWO$_4$, CdWO$_4$, CaF, and TeO$_2$) and, thus, cannot be convincingly described by the model of \cite{probst_model_1995}: Any particle interaction initially creates non-thermal phonons, i.e.~phonons exceeding thermal energies (at ~\unit[10]{mK} operating temperature). These high-frequency phonons may be directly absorbed in the thermometer (non-thermal component) or first thermalize in the absorber before being registered (thermal component), for details the reader is referred to  \cite{angloher_results_2017,probst_model_1995}. We repeated the pulse-shape analysis with data from the second prototype, the result is depicted in figures \ref{fig:2comp} and \ref{fig:3comp}.

The black data points in the top windows show a standard pulse created by averaging $\mathcal{O}(100)$ events of the same energy deposition. We fit this average pulse two times, in figure \ref{fig:2comp} with the model of  \cite{probst_model_1995} consisting of a non-thermal and a thermal component. Adding a second \textit{thermal} component to the model leads to the fit result shown in figure \ref{fig:3comp}. Looking at the residuals of both fits evidently shows that the additional, third component significantly improves the fit result and almost perfectly describes the measured pulse, whereas the two-component fit does not. The peculiarity of NaI is a long decay time of the pulse. In the two-component fit this behaviour dominates the fit leading to a lack of thermal component in the main peak and, thus, to considerably high residuals. The three-component fit, instead, attributes the long decay to the third component, and the primary thermal component to mainly contribute in the range of the main peak. As the decay times exceed the maximally available record length of the present DAQ we will switch to a dead-time free DAQ in the future, thus allowing to record the full tail which is a immensely useful tool to understand the origin of this third pulse component.

\section{Gamma Calibration Data}

Figure \ref{fig:LY} shows data from the measurement of the second prototype in the light yield versus energy plane. The light yield is defined as the ratio of light to phonon signal, set to one by definition for $\beta/\gamma$-events. The black solid lines indicate a fit of the $\beta/\gamma$-band to the data with \unit[80]{\%} of the respective events expected inside. We do not observe a non-proportionality of the light output at low energies, as it is common for other inorganic scintillators, for instance CaWO$_4$ \cite{lang_scintillator_2009}. Instead, the mean line of the $\beta/\gamma$-band is well modeled by a simple linear function, the width of the band includes contributions from the baseline resolutions of phonon and light detector, a Poisson-statistics term accounting for the quantization of the scintillation light photons and a empiric term scaling with light energy (e.g.~potential position dependencies). Three characteristic features become evident in the $\beta/\gamma$-band, the line at \unit[$\sim$60]{keV} originating from gammas from the $^{241}$Am source, so-called escape peaks from I at \unit[$\sim$30]{keV} \cite{angloher_csi_2016,angloher_results_2017} and a line at around \unit[3]{keV} caused by $^{40}$K Auger electrons.\footnote{This line is below the phonon trigger threshold, but is triggered by the light detector.} Potassium is the main and probably toughest contaminant to fight against for NaI-based DM searches due to its chemical affinity with sodium.

To calculate the nuclear recoil bands for recoils off Na (blue) and I (green) we use the quenching factors (QFs), which quantify the reduction in light output for a certain event type compared to a $\beta/\gamma$-event of the same energy, of \cite{tretyak_semi-empirical_2010}. We want to note that these QFs are determined for Tl-doped NaI at room temperature, thus, deviations thereof cannot be excluded for the use in COSINUS. To overcome this uncertainty we foresee measurements of the QFs at a neutron beam facility previously used by the CRESST experiment \cite{strauss_energy-dependent_2014}. 

The magenta band marks the region expected for events with no light signal. This band is highly populated; the events therein originate from the carrier crystal. Those carrier events have a different pulse-shape than events in the NaI absorber (in particular they do not have the long decay times as discussed in section \ref{sec:pulse-shape}), however the signal-to-noise ratio is not good enough to allow for a discrimination down to threshold. Therefore, no cut was applied.% The number of events in this carrier band by far exceeds the expected particle count rate, it is also substantially higher than the rate observed in \cite{angloher_results_2017}. We identified the switch from silicon oil for the absorber-carrier-interface, as used in \cite{angloher_results_2017}, to epoxy resin as reason for the increase. The resin lead to micro-cracks of the carrier, confirmed afterwards by optical inspection.

We operated the phonon detector with a hardware trigger threshold of \unit[($8.26\pm0.02$(stat.))]{keV}, which is still significantly above the COSINUS design goal of \unit[1]{keV}. We achieve a baseline resolution for the phonon detector of \unit[$\sigma=1.01$]{keV} increasing to \unit[$\sigma=4.5$]{keV} at the $^{241}$Am-peak at \unit[60]{keV}. This comparatively high dependence of the resolution on the energy is another aspect to be studied in the future to reach a full understanding of the phonon properties of NaI.

\section{Absolute Light Yield} \label{sec:LY}

The absolute light yield denotes the fraction of the energy deposited in the NaI crystal which is measured by the light detector. To determine this quantity we compare the light pulse amplitude of \unit[60]{keV} energy depositions in the NaI originating from $^{241}$Am to direct hits of \unit[5.9] and \unit[6.5]{keV} X-rays originating from an $^{55}$Fe source dedicatedly installed to solely irradiate the light detector. To account for pulse-shape differences between scintillation light events and direct hits we weigh the respective amplitudes by the integral over the corresponding light pulses. For this measurement we achieve an absolute light yield of \unit[13.1]{\%}, which corresponds to an average of \unit[$\sim$40]{photons/keV} given a mean scintillation photon energy of \unit[3.3]{eV}. This value is roughly three times higher than for the first prototype \cite{angloher_results_2017} and achieved by replacing the wafer-like light absorber by a beaker-shaped one convincingly showing the potential of this light detector technology.

\section{Conclusions and Outlook}

Operating NaI as cryogenic calorimeter is a challenging task. After the first feasibility measurement outlined in \cite{angloher_results_2017} this paper contains the proof-of-principle measurement of the final detector design. We observe a peculiar phonon pulse-shape previously not seen in numerous other materials. Understanding it is not only highly significant in the light of dark matter sensitivity for COSINUS, but also bears relevant results in the field of low-temperature solid-state physics. 
The main goal of this measurement was to test a module in the final configuration featuring a beaker-shaped light detector. This goal is not only achieved, but we measure \unit[13.1]{\%} (\unit[$\sim$40]{photons/keV}) of the energy deposited in the NaI as scintillation light exceeding the original COSINUS design goal (\unit[4]{\%}) by more than a factor of three. This proves that undoped NaI exhibits an outstandingly high light output at low temperatures, unmatched by any other NaI-based dark matter search which are all based on Tl-doped NaI crystals operated at room temperature (ANAIS e.g. reports a value of \unit[15]{photoelectrons/keV} \cite{coarasa_annual_2017}). 

The (unfortunate) properties of NaI (high Debye temperature, hygroscopicity, low hardness etc.) make the development of a NaI-based cryogenic calorimeter a stony path. With the measurement of the second prototype, as discussed in this paper, we are advancing on this path and prove that we cleared the rock of absolute light output out of our way towards a COSINUS dark matter detector.

\section*{Acknowledgements}
\footnotesize{This work was carried out in the frame of the COSINUS R\&D project funded by the INFN (CSN5). We want to thank the LNGS mechanical workshop team E. Tatananni, A. Rotilio, A. Corsi, and B. Romualdi for continuous help in the overall set-up construction and M. Guetti for his cryogenic expertise and his constant support.}\newline

\section*{References}
%\begin{thebibliography}{9}
%\bibitem{iopartnum} IOP Publishing is to grateful Mark A Caprio, Center for Theoretical Physics, Yale University, for permission to include the {\tt iopart-num} \BibTeX package (version 2.0, December 21, 2006) with  this documentation. Updates and new releases of {\tt iopart-num} can be found on \verb"www.ctan.org" (CTAN). 
%\end{thebibliography}

\bibliographystyle{h-physrev}

\bibliography{main}

\end{document}